\documentclass[fp,twocolumn]{jpsj3_nohead}
\usepackage{txfonts}
\usepackage{color,graphicx}

\title{Magnetic Nonlinear Response of UPt$_3$: An augmented Landau approach}

\author{Trevor D. Ford$^1$, K. Matsumoto$^2$,  and B.S. Shivaram$^{1,\dagger} $}
\inst{$^1$Department of Physics, University of Virginia, Charlottesville, VA. 22901\\

$^2$Hokkaido University of Education - Asahikawa, Hokumon-cho 9, Asahikawa, 070-8621, Japan}

\abst{Several heavy fermion materials, including UPt$_3$, exhibit a rapid but gradual rise in the magnetization at a critical field, without an apparent phase transition at any temperature $T>0$, with the possibility of a first order transition at $T \equiv0$. To model such a quantum phase transition it is most appropriate to develop approaches considering the quantum nature of the spins.  Within a fully classical framework, we show that it is sufficient to start from a Landau-type free energy with an added Bragg-Williams entropy term to arrive at a number of key experimental features as seen in UPt$_3$.  In particular, we show that correctly arriving at the measured (low-field) higher order susceptibilities necessarily invokes an isobestic (crossing) point at a high field in the magnetization isotherms.  We also present a full analysis of the angular dependence of the (low-field) linear and nonlinear susceptibilities which when extended also capture the anisotropic high field response of the magnetization. Key to this success is the proper conversion of the evaluated magnetization from constant volume to a constant pressure situation relevant at high fields in heavy fermion materials.}

\begin{document}
\maketitle

\section{Introduction}

A characteristic behavior of many heavy fermion magnets is that they undergo a significant change in magnetization upon a small change in external magnetic field at a critical field. Additionally, the magnetization of these systems tends to saturate at field strengths much greater than the critical field strength. %Any proposed model for metamagnetism must first and foremost be capable of reproducing these behaviors. - obvious statement.%
In these heavy fermion systems it is also observed that the magnetization takes roughly the same value at a critical field, $h_c$, for all temperatures below some characteristic temperature, $\approx T_m $. Thus, there is a “crossing-point” in the magnetization isotherms, for $T < T_m$,  which is clearly seen in UPt$_3$ \cite{FifthOrderPRB2014} and Sr$_3$Ru$_2$O$_7$ \cite{ShivaramSRO2018}. Such crossing points, termed isobestic points, are a more general phenomenon encountered frequently in strongly correlated electronic systems \cite{VollhardtPRB2013, VollhardtPRL1997}.\\

Yet another characteristic of these heavy fermion systems, including UPt$_3$ and CeRu$_2$Si$_2$, is the occurance of a broad maximum in the linear susceptibility\cite{FringsJMMM1983, FlouquestJMMM1985}, $\chi_1$,  at a particular temperature, $T_1$.  As reported by one of us\cite{UniversalityPRB2014} the leading order nonlinear susceptibility, $\chi_3$, in UPt$_3$ also exhibits a maximum at a temperature $T_3$ close to $0.5T_1$. Furthermore, it was shown in this work that this correlation between $T_1$ and $T_3$ is found in many strongly correlated systems where the third order susceptibility has been measured.  A similar scaling between $T_1$ and $h_c$, has also been known for a while\cite{Hirose2011}.  A single energy scale model involving quantum spins was introduced\cite{UniversalityPRB2014} and further elaborated to capture these observed correlations\cite{CelliPhilMag2020}.  \\

In analyzing the magnetic response in CeRu$_2$Si$_2$, Matsumoto and Murayama recently introduced a model with a Landau-type free energy appended with a Bragg-Williams \cite{Murayama3} entropy term. The Bragg-Williams entropy is the entropy of a spin-$\frac{1}{2}$ Ising system expressed in terms of the system's magnetization.\cite{Lubensky} This entropy term can be used as an appoximation for the entropy of a collection of local magnetic moments in an external field because the normalized spontaneous magnetization of such a system does not vary appreciably with respect to the spin, $S$.\cite{Yamauchi} Including the Bragg-Williams entropy term allows the model to better capture the behavior of materials where localized spins contribute to the magnetic response, such as CeRu$_2$Si$_2$ or UPt$_3$. Matsumoto's choice of the free energy is given below: 

\begin{align}
    F\left[m,h\right]=\frac{a}{2}m^2-hm-chm^3-\left(\frac{T}{k_BT_m}\right)S_{BW},
    \\
    S_{BW}=-k_B\left(\frac{1+m}{2}\ln{\frac{1+m}{2}}+\frac{1-m}{2}\ln{\frac{1-m}{2}}\right).
\end{align}

In their analysis, they employed the fact that the linear susceptibility in CeRu$_2$Si$_2$ remains roughly constant for temperatures below $T_m$.  Therefore they allowed the parameter $a$ in the equation above to yield a magnetization that counteracts the high temperature Curie-Weiss behavior and thus yields a temperature independent linear susceptibility at low $T$. Further, the parameter $c$ was set to a temperature independent constant value to ensure that the differential susceptibility $\frac{dm}{dh}$ possesses a maximum at the required critical field, $h_c \approx 8$ T. %The choice of the Bragg-Williams entropy term is appropriate since a non-ordering metamagnet may be viewed as a mixture of up and down spins that are constantly fluctuating.
\\

However, with the choice of $a(T)$ made by Matsumoto and Murayama the occurrance of the maximum in $\chi_1$ and correlations between $T_1, T_3$ and the critical field, $h_c$ are ignored. In this paper we extend their phenomenological mean-field model to account for the peaks in the susceptibiliies and the correlations and match the model results to our experimental data for UPt$_3$. We also present a full analysis of the angular dependence of the (low-field) linear and nonlinear susceptibilities.  The same analysis extended to high fields also accounts for the anisotropic high field response of the magnetization.\\

An integral part of our analysis is the recognition that heavy fermion magnets often demonstrate strong magnetostriction. Since model calculations always start with the assumption of constant volume conditions it is imperative that magnetostriction be accounted for and a transormation from constant volume to constant pressure be applied. This was carried out elegantly by Matsumoto and Murayama \cite{Murayama3} and we employ the same technique for our present work. This conversion however, is not necessary for the analysis of low-field behaviors since magnetostriction is negligible in this limit. \\

%In addition to maxima in the low-field susceptibilities, our data for UPt$_3$ also shows the “cross-over” of many $m-h$ curves for temperatures below the metamagnetic transition temperature as commonly expected of metamagnetic HFCs. In previous publications, the parameter $c$ was used to enforce a matching of the critical fields between theory and experiment at zero temperature. However, choosing $c$ to be constant limits the behaviors one can achieve with the model. Namely, one can enforce the “cross-over” behavior observed in our experimental data and the data from many other HFC metamagnets if $c$ is instead allowed to vary with temperature. Thus, in this paper we also aim to propose a form for $c(T)$ that enforces a “cross-over” at the critical field.\\

\section{Low-Field Susceptibilities}

The free energy functional of Eq. (1) when minimized yields the equilibrium $m-h$ isotherms. Thus, it is straightforward to verify that the following relationship between $h$ and $m$ must hold:

\begin{equation}
   \frac{\partial F\left[m,h\right]}{\partial m}=am-h-3chm^2+\frac{1}{2}\frac{T}{T_m}\log{\frac{1+m}{1-m}}=0.
\end{equation}
  
From this constitutive relation we can derive the leading terms of the first-, third-, and fifth-order low-field susceptibilities and the relationships between these susceptibilities as shown below:

\begin{equation}
    \chi_1=\frac{T_m}{T+aT_m},
\end{equation}
\begin{equation}
    \chi_3=\frac{3cT_m^3}{(T+aT_m)^3}=3c\chi_1^3,
\end{equation}
\begin{equation}
    \chi_5=\frac{18c^2T_m^5}{(T+aT_m)^5}=18c^2\chi_1^5.
\end{equation}

The full forms of $\chi_3$ and $\chi_5$ involve additional correction terms, but these terms are small in the regime of data collected for this study relative to the size of the leading order terms. Thus, the correction terms can be safely neglected.

As per Eq. (4) a peak in the first-order susceptibility will occur if $T+aT_m$ has a minimum for some $T>0$ i.e.:

\begin{equation}
   \left.\frac{da}{dT}\right|_{T=T_1}=-\frac{1}{T_m}.
\end{equation}

One form for $a(T)$ that can satisfy this condition and thereby produce a maximum in the first-order susceptibility is a simple exponential with a constant offset,

\begin{equation}
    a(T)=Ae^{-kT}+B,
\end{equation}

\begin{equation}
    \chi_1\approx\frac{T_m}{T+BT_m}.
\end{equation}

At this point, we do not make any claims as to the physical origin of this exponential dependence. The three tunable parameters, $A$, $B$, and $k$, can be adjusted to produce a linear susceptibility that best matches experimental data over the available temperature range (2K - 300K). In the high temperature limit we obtain the Curie-Weiss behavior: 

\begin{equation}
    \chi_1\approx\frac{T_m}{T+BT_m}.
\end{equation}

Thus the Curie constant is $C=T_m$ and the Curie-Weiss temperature is $\Theta_{CW}=BT_m$. The relationship between the parameters $A$, $k$ and $T_1$ is given as:

\begin{equation}
   A=\frac{e^{kT_1^*}}{kT_m}.
\end{equation}

Combining both the Curie-Weiss relationships and the relationship of Eq. (10) yields a model for $a(T)$ that only explicitly requires a univariate least-squares fitting for the parameter $k$. However, we found that using the parameter values determined via such a univariate fit as an initial guess in a multivariate fit with the full form of Eq. (8) produced better agreement with the data. This approach was used for fitting data measured for the UPt$_3$ sample along the $a$ axis, with the fit given in Fig. 1 top part. The $c$ axis data does not exhibit a maximum in $\chi_1$, so the initial guess for the parameter $A$, which depends on $T_1$, in the multivariate fitting was simply set to unity. The fit to the $c$ axis data is given in the bottom panel of Fig. 1.\\

\begin{figure}
\centering
    \includegraphics[scale=0.5]{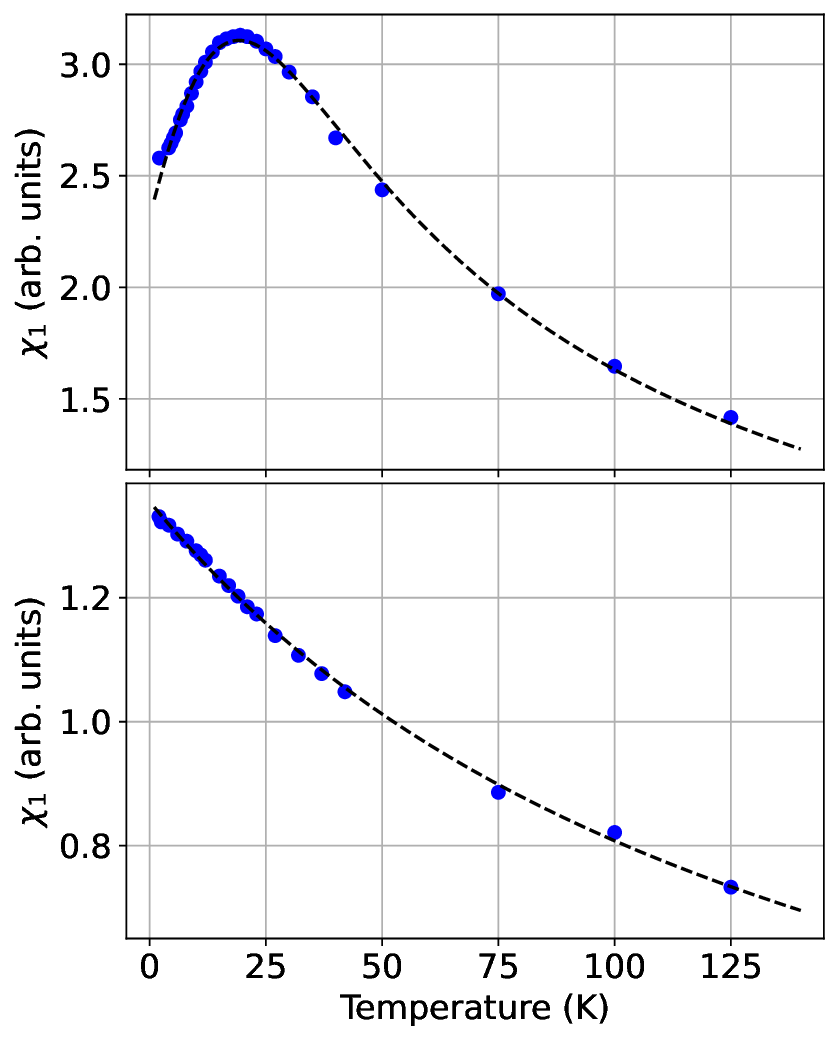}
\caption{Model fit to experimentally measured $\chi_1$ data along the $a$ axis (top panel) and the c-axis (bottom panel) as a function of temperature. To ensure a reasonable fit, an initial parameter guess for the multivariate fitting procedure was provided by a fit to the high temperature data according to the Curie-Weiss law. For the data in the top panel, the fit parameters are $A=0.2465$, $B=0.1856$, $T_m=234$ K, and $k=7.89\times 10^{-2}$ K$^{-1}$; for the data in the bottom panel, the fit parameters are $A=0.3887$, $B=0.3491$, $T_m=200$ K, and $k=4.54\times 10^{-8}$ K$^{-1}$.}
\label{f1}
\end{figure}

\begin{figure}
\centering
    \includegraphics[scale=0.5]{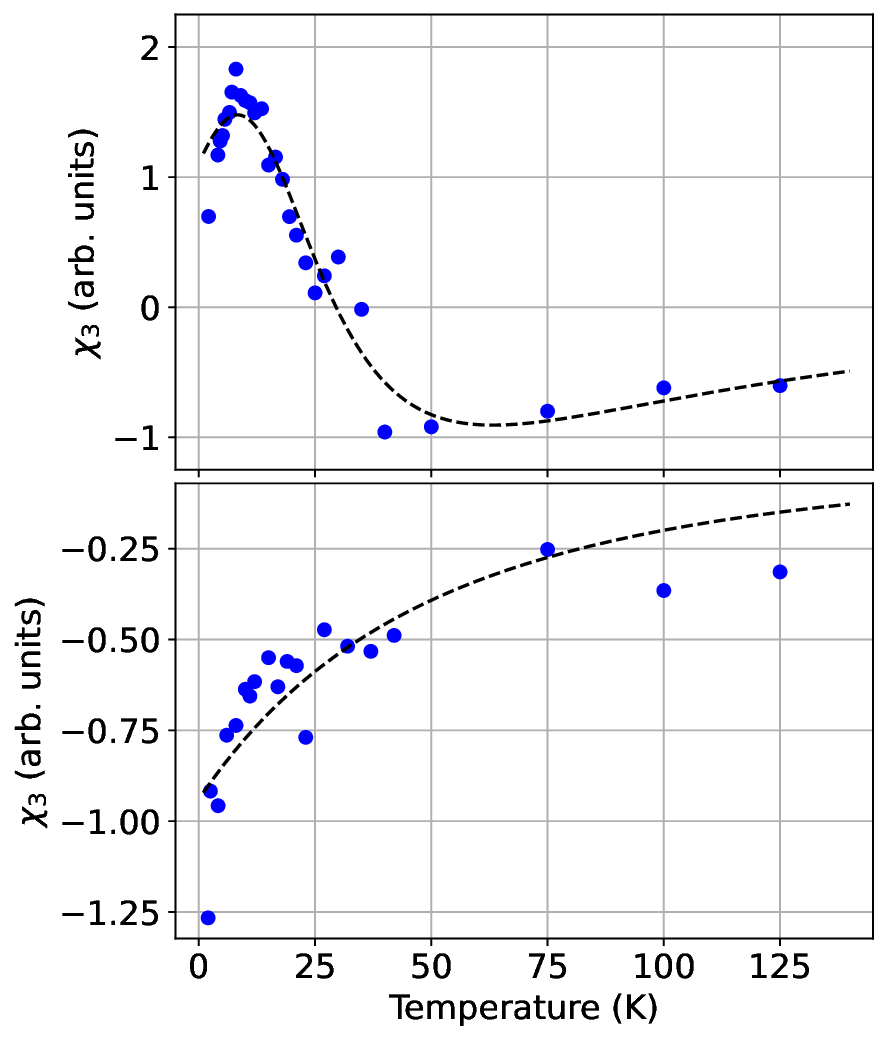}
\caption{Model fit to experimentally measured $\chi_3$ data along the $a$ axis (top panel) and $c$ axis (bottom panel) as a function of temperature.  To ensure a reasonable fit, an initial parameter guess for the multivariate fitting procedure was provided by a fit to the high-T data according to the Curie-Weiss law. For the data in the top panel, the fit parameters in the model for $c(T)$ are $A=0.213$, $B=-0.183$, and $k=5.10\times 10^{-3}$ K$^{-1}$; for the data in the bottom panel, the fit parameters are $A=-5$, $B=-0.1259$, and $k=20$ K$^{-1}$.}
\label{f2}
\end{figure}

In their approach Matsumoto and Murayama set the parameter $c$ to be constant with respect to temperature. Given the relations of Eqs. (4) - (6), it is apparent that such a choice would force the higher order susceptibilities to peak at roughly the same temperature as $\chi_1$; the correction terms contributing to $\chi_3$ and $\chi_5$ only slightly shift this concurrence.  Choosing a constant $c$ necessarily enforces the condition that $T_1\approx T_3\approx T_5$ and as stated earlier however, experimental data for UPt$_3$ shows that $T_3$ is roughly half the value of $T_1$. Thus, to appropriately match the data it is necessary that $c$ must vary with temperature. \\

We propose another simple exponential model for $c(T)$ with the same form as the one for $a(T)$ of Eq. (8). The approach to finding the model parameters for $c$ involves using the relationship between $\chi_3$ and $\chi_1$ in Eq. (5). An exponential fit to the quantity $\frac{\chi_3}{3\chi_1^3}$ will give the model parameters for $c$. In the context of the current available dataset however, we could not compute this quantity for a sufficiently large number of temperatures to ensure an appropriate fit. Instead, our approach was to fit the form of $c$ to $\chi_3$ data based on Eq. (5).  Since the parameter $T_m$ is found during the procedure by which $a$ is found, thus only the parameters of the model for $c$ are fit in this process. The results of these fits to the $\chi_3$ data for both the $a$ and $c$ axes are given in Fig. 2, top and bottom respectively.\\

These fitting procedures determine all parameters in the free energy in Eq. (1) for UPt$_3$. Therefore, it is reasonable to suppose that the magnetization versus magnetic field (directed strictly along either the $a$ or $c$ axis) curves at any given temperature can be predicted using these model fits. While it is true that the model has now been determined based on experimental findings, it is presently capable of predicting the magnetization curves under conditions of constant volume. However, the experiments measuring the magnetization of UPt$_3$ samples were performed under conditions of constant pressure. A conversion process between the two types of conditions is necessary for making appropriate magnetization curve predictions to be compared against real experiments.\\

\section{Constant $V$ to Constant $P$ Conversion}

The magnetization versus magnetic field strength curves obtained in experiments are measured under conditions of constant pressure, however the model calculations are always under constant volume conditions. Because the magnetostriction of these materials is oftentimes very significant, it is necessary for our analysis to convert the constant volume results into what they would be at constant pressure. Matsumoto and Murayama provide a recipe for this process, and we summarize the key steps of the conversion here for completeness.

\begin{figure}[h]
\vspace{0cm}
\centering
    \includegraphics[scale=0.57]{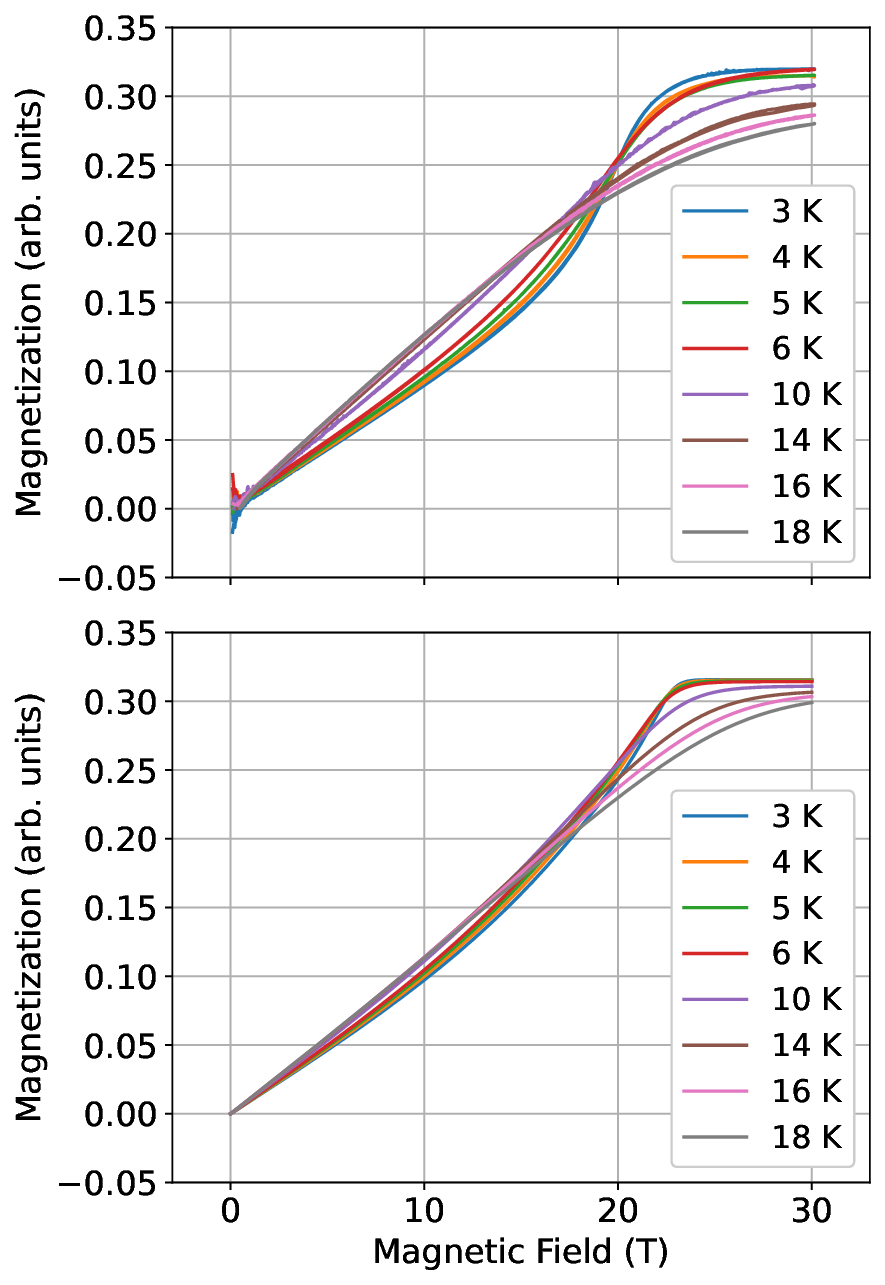}

\caption{The top plot shows the experimentally measured magnetization of the UPt$_3$ sample as a function of the applied external magnetic field for the temperatures specified in the legend. The data was collected with the magnetic field directed along the $a$ axis of UPt$_3$. The bottom plot shows the model predictions for those same experimental measurements in the top panel. The model predictions have been scaled to ensure that both magnetizations approach $0.32$ in the limit of large applied field.}
\label{f3}
\end{figure}

The conversion between isotherms at constant volume to isotherms of constant pressure begins with a calculation of the magnetostriction using the calculated constant volume susceptibility, and experimentally determined parameters such as the isothermal compressibility, $\kappa_T$, the Gruneisen parameter, $\Gamma$, and the molar volume, $V_0$. The values of these parameters in 
the case of UPt$_3$ are $\kappa_T=4.8\cdot 10^{-12}$ Pa$^{-1}$,\cite{vanSprang} $\Gamma=73$,\cite{FRANSE1988147} and $V_0=4.223\cdot 10^{-5}$ m$^3$/mol.\cite{RevModPhys.74.235} Magnetostriction describes the amount of constriction (or expansion) of a material when an external magnetic field is introduced. A large magnetoscriction noticably alters the susceptibility of a material between experiments at constant pressure and constant volume. Since many heavy fermion magnets including UPt$_3$ demonstrate a strong magnetostrictive response, the calculation of mangetostriction as given below is vital for proper comparison:  
\begin{equation}
\frac{\Delta V}{V_0}=\kappa_TN_Ak_BT_m(V_0)\frac{\Gamma}{V_0}\int_0^{h}h'\chi^{V}(h')dh'.
\end{equation}

The remaining steps of the conversion to constant pressure conditions are straightforward, and involve direct substitution of the magnetostriction into the following expressions:
\begin{equation}
R\left(\Delta V\right)=\left(1+\frac{\Delta V}{V_0}\right)^{\Gamma},
\end{equation}
\begin{equation}
\chi^{P_0}(h)=R\left(\Delta V\right)\chi^{V_0}(R\left(\Delta V\right)h).
\end{equation}

This process converts the constant volume susceptibility into a constant pressure susceptibility. It's important to note that this conversion process does not change the zero-field values of $\chi_1$ and $\chi_3$, thus the fitting procedures for $a(T)$ and $c(T)$ in the previous section are still valid. Integrating the susceptibility yields the constant pressure $m-h$ isotherms.  These model calculations along with the experimental isotherms for UPt$_3$ are shown in Fig. 3.\\

\section{Anisotropy}

UPt$_3$ belongs to the hexagonal crystal system and has $D_{6h}$ symmetry. Considering a general expansion of free energy in terms of even powers of $H$, Eq. (15), and enforcing the symmetries of $D_{6h}$ onto the (fully symmetric) susceptibility tensor will cause a majority of the tensor components to vanish.

\begin{equation}
F=\frac{1}{2!}\chi^{ij}h_ih_j+\frac{1}{4!}\chi^{ijk\ell}h_ih_jh_kh_{\ell}+\cdot\cdot\cdot
\end{equation}

The point group $D_{6h}$ possesses a six-fold rotation axis, which we will orient along the z-axis without loss of generality. This rotation is represented in typical fashion as a 3-by-3 rotation matrix, $R^i_j$. The transformations of the second- and fourth-order tensors under this rotation are:

\begin{equation}
\Tilde{\chi}^{rs}=R^r_iR^s_j\chi^{ij},
\end{equation}

\begin{equation}
\Tilde{\chi}^{rstu}=R^r_iR^s_jR^t_kR^u_\ell\chi^{ijk\ell}.
\end{equation}

The rotation matrix for a six-fold rotation about the $z$-axis is:

\begin{equation}
    R=\begin{bmatrix}
        \frac{1}{2} & -\frac{\sqrt{3}}{2} & 0 \\
        \frac{\sqrt{3}}{2} & \frac{1}{2} & 0 \\
        0 & 0 & 1
    \end{bmatrix}.
\end{equation}

Solving the equations in which the transformed tensors are equivalent to the original tensors indicates which components of the tensors must vanish, and also gives the relationships between any non-vanishing components. \\

For the rank two susceptibility, it was found that the only non-vanishing components are $\chi^{11}=\chi^{22}$ and $\chi^{33}$. This suggests that the second-order contribution to the free energy is of the form $\frac{1}{2!}(\chi_{a}\sin^2\theta+\chi_{b}\cos^2\theta)h^2$, where $\chi_{a}$ and $\chi_b$ are linear combinations of the non-vanishing components of the tensor and $\theta$ is the angle between the applied magnetic field and $a$ axis. The specific linear combinations encoded in these $\chi_a$ and $\chi_b$ are not necessary for the current discussion. For the rank four susceptibility, it was found that the only non-vanishing components are $\chi^{1133}=\chi^{2233}$ (and index permutations) and $\chi^{3333}$. Thus, the fourth-order contribution to the free energy is of the form $\frac{1}{4!}(\chi_{c}\cos^4\theta+\chi_{d}\cos^2\theta)h^4$, where again $\chi_{c}$ and $\chi_{d}$ are some linear combination of the non-vanishing components of the tensor whose specific form is not necessary for this discussion. \\

The magnetization can be found from the free energy as:

\begin{equation}
    m=(\chi_{a}\sin^2\theta+\chi_{b}\cos^2\theta)h+\frac{1}{6}(\chi_{c}\cos^4\theta+\chi_{d}\cos^2\theta)h^3+\cdot\cdot\cdot
\end{equation}

The first term corresponds directly to $\chi_1$ while the third term corresponds directly to 
$\chi_3$. This suggests that if we measure $\chi_1$ and $\chi_3$ at two different angles $\theta$ with respect to the z-axis, say $0$ and an angle slightly less than $\frac{\pi}{2}$, then we can solve for the values of $\chi_a$, $\chi_b$, $\chi_c$, and $\chi_d$ thus enabling us to arrive at the full angular dependence of both the linear and the third order susceptibilities. A choice of $\theta=\frac{\pi}{2}$ for one of the angles will have to be avoided since according to Eq. (19) $\chi_3$ vanishes at that angle thus restricting the information needed. The results of such a model prediction of the angular dependence of $\chi_1$ and $\chi_3$ for $D_{6h}$ (hexagonal symmetry) are shown in Fig. 4. The experimentally observed behavior of $\chi_3$ for the two cases, field parallel to c-axis and perpendicular to c-axis, as shown in Fig. 2 is clearly reproduced.  A similar anisotropy for $\chi_3$ can also be seen in the data for CeRu$_2$Si$_2$ \cite{ParkJPCM1994}, a tetragonal system.  The full anisotropy of $\chi_3$, including measurements at intermediate angles has also been studied experimentally in another tetragonal system URu$_2$Si$_2$ \cite{TrinhPRL2016}.  Here symmetry dictates that there be three independent components of the third order susceptibility tensor.  However, it is found experimentally that only the component with the $\cos^4(\theta)$ dependence survives, thus providing additional insights into the magnetism of URu$_2$Si$_2$ \cite{ChandraPhysica2018}. It would be useful to perform such full angular dependent studies of $\chi_3$ in UPt$_3$ also.  Additional motivation for such studies comes from the unique behavior of the anisotropic magnetostriction in UPt$_3$ \cite{deVisser} which implies that the magnetostriction is identically zero along the c-axis for an angle of $51\deg $ away from the c-axis whereas it is significant along the basal plane.  As shown through ultrasound measurements a tricritical point occurs precisely at this angle at the high field of 31 T \cite{ShivaramSR2019}.  However the vanishing of the magnetostriction along the c-axis when the field is oriented at this special angle occurs at all fields.  It would be interesting therefore to look for deviations of the observed behavior of $\chi_3$ (and $\chi_1$) from predictions of Eq. (19) in this angular range.\\

\begin{figure}[h]
\vspace{0cm}
\includegraphics[scale=0.5]{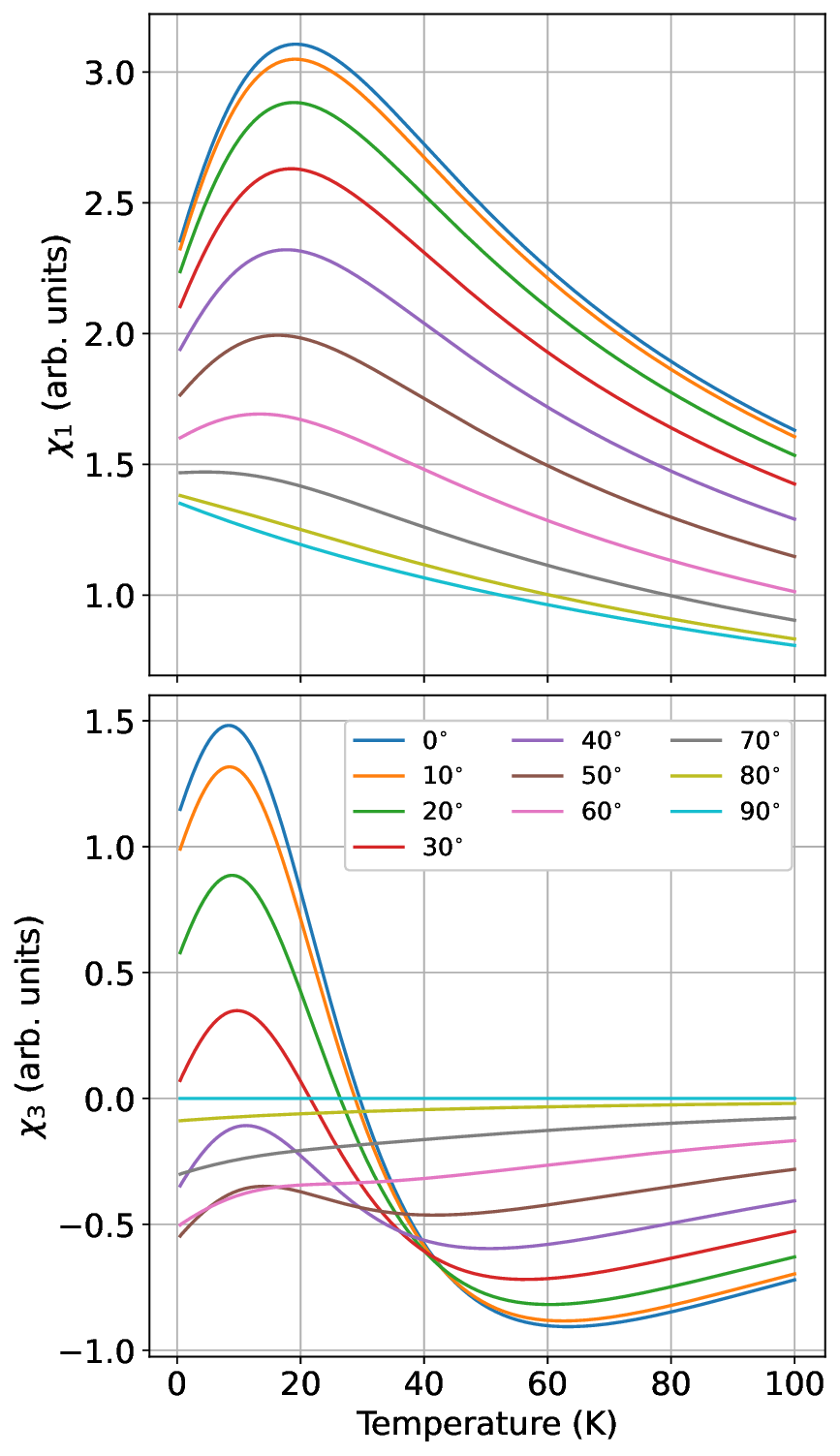}
\caption{Predictions of the first order low-field susceptibility as a function of temperature for various crystal orientations relative to the external magnetic field.}
\label{f4}
\end{figure}

\begin{figure}
\includegraphics[scale=0.5]{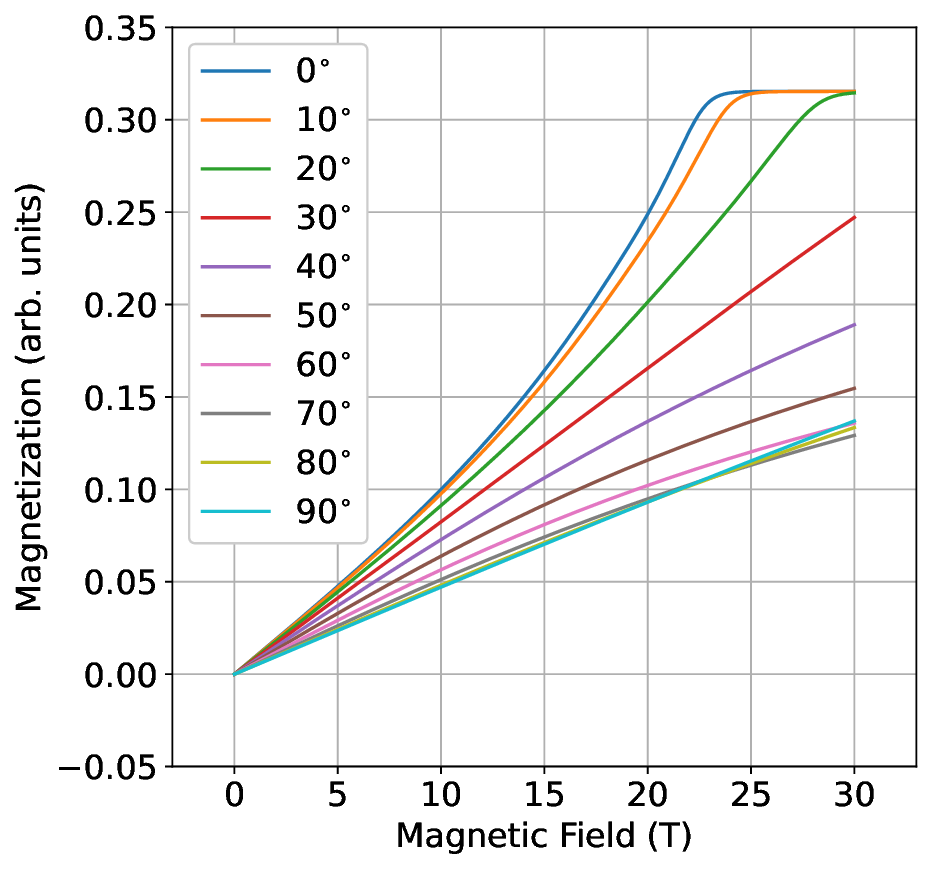}
\caption{The magnetization isotherms as obtained in the model for various orientations of the field with respect to the a-axis.  The point of inflection (i.e. the critical field) moves to larger values much more rapidly than seen in experiments.}
\label{f5}
\end{figure}

Since we now have the temperature dependence of $\chi_1$ and $\chi_3$ for arbitrary angle we can find model fits for $a(T)$ and $c(T)$ for any given angle using the procedure detailed earlier in this paper.  Thus the parameters $a$ and $c$ are effectively turned into $a(T,\theta)$ and $c(T,\theta)$. These parameters can be used in Eq. (3) to solve for the full $m-h$ curves at any desired $T$ and $\theta$, and a subsequent $V$-to-$P$ conversion will predict the expected measured magnetization, as shown in Fig. 5.  From such $m-h$ curves for any given $T$ and $\theta$, we can obtain a critical field, defined as the point of inflection in the model isotherms.  The angle dependent model critical field is shown in Fig. 6 (solid line - left vertical axis), and it follows a clear $\sec(\theta)$ dependence as found experimentally \cite{SuslovJLTP2000, SuslovIJMP2002}.  The differential susceptibility which peaks at the inflection point can also be found as a function of angle, dotted line (right vertical axis) in Fig. 6, and has a clear $\cos^2(\theta)$ dependence also seen in experiments \cite{SuslovJLTP2000}.  However, since our approach here has beeen purely clasical it is not surprising that there are significant quantitative discrepancies. The critical field quickly diverges to very large values with a tilt away from the basal plane more rapidly than the experimental values. The differential susceptibility is also much weaker compared to experiments thus pointing to the importance of including quantum fluctuations in a more complete treatment of magnetism in these systems.

\begin{figure}
\includegraphics[scale=0.5]{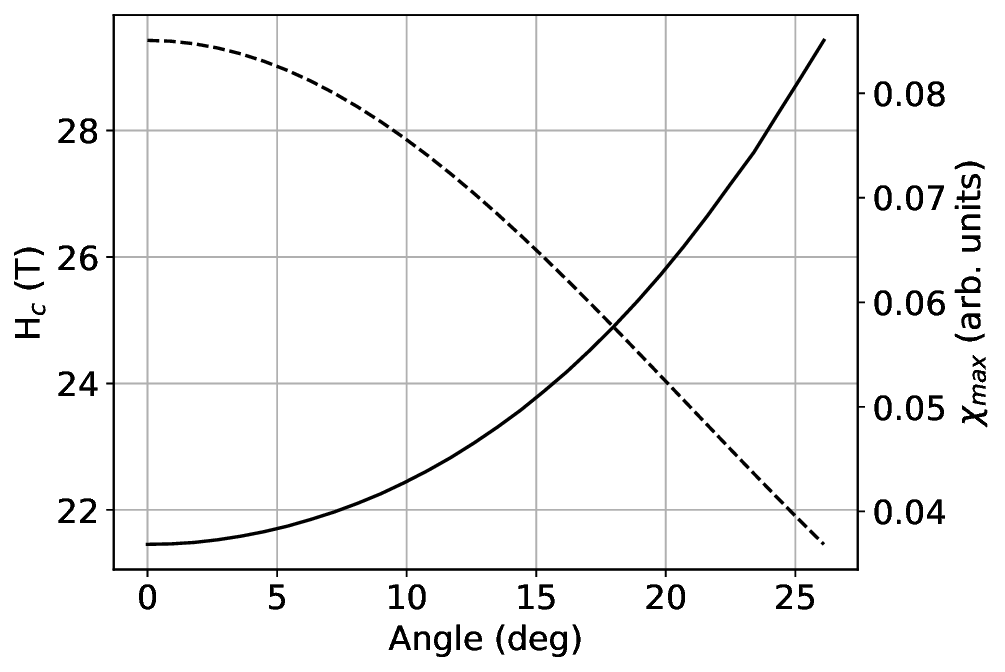}
\caption{Shows the critical field, $h_c$ obtained from model isotherms (solid line - left vertical axis)  as in Fig. 5 and the corresponding values of the differential susceptibility at $h_c$ (dashed line - right vertical axis). The solid line follows a $\sec(\theta)$ behavior and the dashed line obeys $\cos^2(\theta)$ as seen in experiments. \cite{SuslovJLTP2000} }
\label{f6}
\end{figure}

\section{Discussion}  In this work we have presented an approach to understand magnetism in heavy fermion systems such as UPt$_3$ which exhibit a smooth crossover in the magnetization with field even at the lowest temperatures. In particular we have modeled accurately the measured linear and nonlinear susceptibilities within a purely classical approach and demonstrated that an ``isobestic" or crossing point at high fields arises naturally. Our model captures qualitatively several other key observed features as discussed in the text.  Nevertheless, and not surprisingly, noticeable discrepancies exist. In the model the critical field diverges strongly with angle and with correspondingly weak values of the differential susceptibility at $h_c$ -  results that are at variance with experiments.  In addition, experimentally it is also found that the inverse peak value of $\frac{dm}{dh}$ as a function of temperature is linear over a very significant temperature range in both UPt$_3$ - see Fig. S4 in supplementary information and CeRu$_2$Si$_2$ \cite{FlouquetPhysica1995}.  In addition, in CeRu$_2$Si$_2$ it has been shown that the deviation of magnetization away from its value at the crossing point, $\Delta M$, varies as $T^2$ in the low temperature limit. The origin of the $T^2$ dependence comes from fermionic spins and is not captured in our model (see supplementary information Fig. S1). These discrepancies could possibly be set right without departing too much from our current approach.  Quantum spins can be introduced leaving the free energy functional form unaffected. Quantum effects will affect the amplitude of each term's coefficient in the free energy. Currently the Bragg-Williams entropy form is for the spin-$\frac{1}{2}$ Ising model. It could be replaced by an appropriate expression for the quantum spins. The amplitude of the parameter $a(T)$ and $c(T)$ can also be changed in models that go beyond mean-field.

Since the ``seeds" of the observed phenomenology are found in a classical model, this suggests that quantum effects merely enhance the phenomenology at low temperatures at times adding a distinct ``quantum imprint." To be sure, over the previous decades there have been several attempts within a microscopic framework to describe heavy fermion magnets \cite{FreericksPRB1992, RaoPhysica1993, SuvasiniPRL1993, SatohPRB2001, BauerPRB2006, YamadaJPSJ2011}.  These attempts while capturing certain features leave open many questions. The occurance of the peaks in the linear and third order susceptibilities and the intrinsic connection of these features to a crossing point in magnetization isotherms and the corresponding scaling relations as illustrated in Figs. 1 and 2 appear never to have been comprehensively demonstrated within a single microscopic model. In fact, even within a fully relativistic first principles band structure based calculation the critical field is overestimated by a factor of two \cite{SuvasiniPRL1993}. Similarly, the angular dependence (Fig. 4) appears to have not beeen explicitly addressed. A more complete evaluation of experimental observables within microscopic theories should help elucidate the precise nature of quantum effects in influencing the crossover in the magnetic response.  We might learn more about quantum fluctuations and phase transitions - are there different flavors of fluctuations that ride on top of classical effects or is there one universal behavior common to all heavy fermions irrespective of structure, dimensionality and composition?\\

\section{Appendix}
We have from Eq. (1) the expression for the free energy functional with the logarithmic term in the Bragg-Williams entropy.  If we keep terms upto fifth order in $m$ in the expansion of the logarithm, we can find all correction terms to $\chi_3$ and $\chi_5$:

$$am-h-3chm^2+\frac{1}{2}\frac{T}{T_m}\left(2m+\frac{2}{3}m^3+\frac{2}{5}m^5...\right)=0.$$

Solving for $m$ yields

$$m=\frac{h}{a-3chm+\frac{T}{T_m}\left(1+\frac{m^2}{3}+\frac{m^4}{5}+...\right)},$$

and an iterative process can be used to express this in terms of powers of $h$.  The forms of the low-field susceptibilties with corrections are:

$$\chi_1=\frac{T_m}{T+aT_m},$$

$$\chi_3=3c\chi_1^3-\frac{T\chi_1^4}{3T_m},$$

$$\chi_5=18c^2\chi_1^5-\frac{\left(20c-1\right)T\chi_1^6}{5T_m}+\frac{T^2\chi_1^7}{9T_m^2}.$$

These corrections contribute significantly less to their respective susceptibilities than the leading terms in the regime of our measured data. Inclusion of the higher order terms with the data measured would not significantly influence the results presented here.

\section{Acknowledgements} The work at the University of Virginia (BSS and TDF) was supported by NSF Award DMR-2016909. The data at high fields presented in this work were obtained at the National High Magnetic Field Laboratory, Tallahassee, Florida supported by the US National Science Foundation. We are grateful to Prof. Pradeep Kumar for many helpful conversations.

%\pagebreak
%\newpage

%\clearpage
%\setcounter{page}{1}

%\includepdf[pages=-,pagecommand={},width=\textwidth]{supplement2.pdf}

\end{document}

\maketitle

\setcounter{figure}{0}
\renewcommand{\figurename}{Fig.}
\renewcommand{\thefigure}{S\arabic{figure}}

\section{Notes on Angular Dependence of Susceptibilities} Enforcing that the tensors possess $D_{6h}$ symmetry requires that the transformed tensor is equivalent to the original tensor, $\Tilde{\chi}=\chi$. For the rank two susceptibility tensor, this condition yields nine linear equations in nine unknowns, while for the rank four susceptibility tensor, the condition instead yields 81 linear equations in 81 unknowns. These systems of equations are homogeneous, so a solution can easily be obtained via finding the row reduced echelon form (RREF) of the coefficient matrix. \\

In developing the coefficient matrix for each system, it is useful to uniquely map tensor components with two or four indices to variables that possess only one label. Each index of a given tensor runs from 1 to 3, so a ternary-to-decimal conversion is appropriate after a relabelling of 1$\longrightarrow$ 0, 2$\longrightarrow$ 1, and 3$\longrightarrow$ 2. As an example, now when the component $\chi^{3213}$ appears in one of the linear equations, its corresponding variable designation will be $\chi_{77}$. Using this mapping will give a specific coefficient matrix which can then be row-reduced. \\

In the following we present additional figures.

\begin{figure}[h]
\centering
    \includegraphics[scale=0.5]{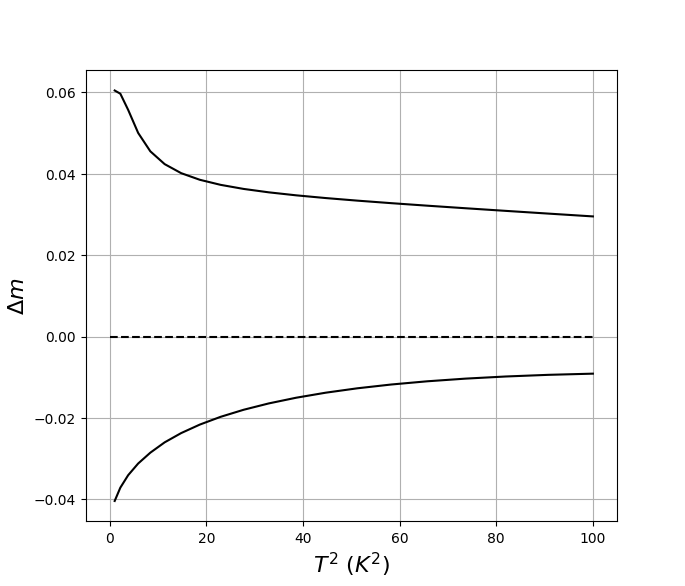}
\caption{A plot of the departure of the magnetization from its "crossing point" value, $\Delta M$ as a function of temperature for different constant fields was presented by C. Paulsen et al., J Low Temp. Phys., \textbf{81}, 317, (1990).  A clear $T^2$ dependence of $\Delta M$ is seen in their data for CeRu$_2$Si$_2$. The figure above shows $\Delta m$ calculated in the model where such a behavior is not seen.  A $T^2$ dependence arises when fermionic spins are involved.}
\label{f2}
\end{figure}

\begin{figure}
\centering
    \includegraphics[scale=0.7]{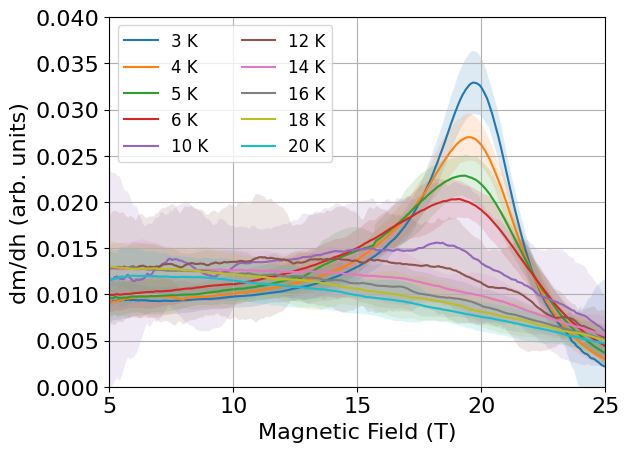}
\caption{Shows the differential susceptibility obtained from the measured magnetization isotherms for UPt$_3$ as a function of magnetic field.}
\label{f2}
\end{figure}

\begin{figure}
\centering
    \includegraphics[scale=0.7]{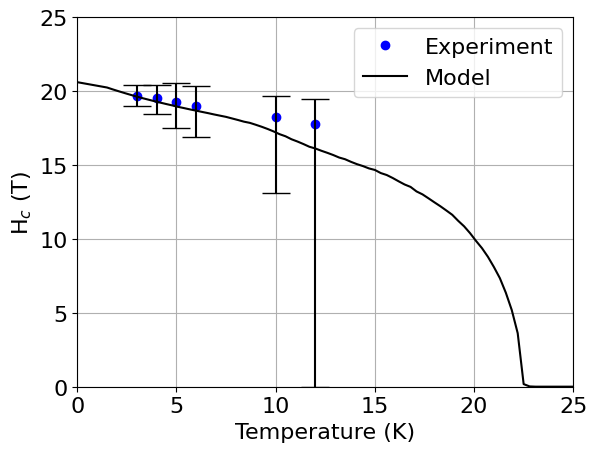}
\caption{The position of the peak values as in fig.2 may be utilised to construct (blue dots) a phase diagram. The solid line shows the behavior expected from the model.}
\label{f2}
\end{figure}

\begin{figure}
\centering
    \includegraphics[scale=0.7]{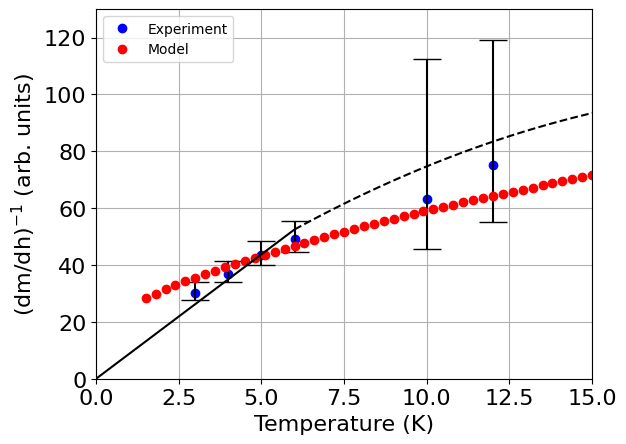}
\caption{The inverse of the experimental differential susceptibility at the critical field follows a linear behavior as a function of temperature at the lowest temperatures (blue dots) passing through zero. A quasilinear behavior is also seen in the model (red dots) but with an apparent non zero intercept at T=0.  A similar linear behavior of the inverse susceptibility is also observed in CeRu$_2$Si$_2$ for T$>$1 K but with a saturation at very low temperatures which gives rise to a non-zero intercept at T=0 (J. Flouquet et al., Physica B, \textbf{215}, 77-87, (1995)).}
\label{f2}
\end{figure}

\begin{figure}
\centering
    \includegraphics[scale=0.7]{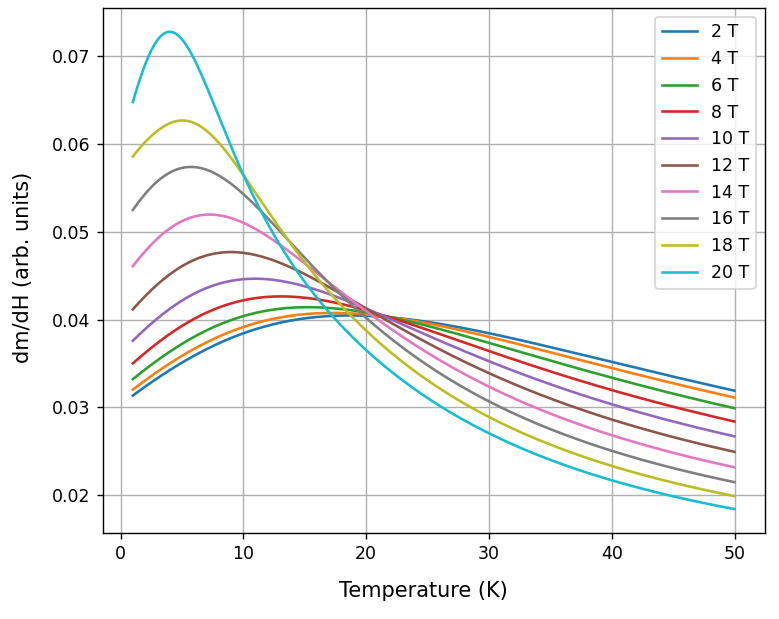}
\caption{While an isobestic point is commonly observed in magnetization isotherms this plot shows that such a point also arises when the differential susceptibility taken at a specific field is plotted as a function of temperature for various fields.}
\label{f2}
\end{figure}